\documentclass[aps, prf, superscriptaddress]{revtex4-2}

\usepackage{amsmath}
\usepackage{amssymb}
\usepackage{graphicx}
\usepackage{hyperref}

\newcommand{\pdiff}[2]{\frac{\partial #1}{\partial #2}}

\newcommand{\rey}{\mbox{Re}}

\newcommand{\mean}[1]{\langle #1 \rangle}

\begin{document}


\title{Wall-Modeled Large-Eddy Simulation Based on Spectral-Element Discretization}

\author{Timofey Mukha}
\email[]{timofey.mukha@kaust.edu.sa, timofey.mukha@protonmail.com}
\affiliation{Department of Engineering Mechanics, KTH Royal Institute of Technology, Stockholm, Sweden}
\affiliation{Computer, Electrical and Mathematical Sciences and Engineering Division, King Abdullah University of Science and Technology (KAUST), 23955-6900 Thuwal,  Kingdom of Saudi Arabia}

\author{Philipp Schlatter}
\email[]{pschlatt@mech.kth.se, philipp.schlatter@fau.de}
\affiliation{Department of Engineering Mechanics, KTH Royal Institute of Technology, Stockholm, Sweden}
\affiliation{Institute of Fluid Mechanics (LSTM), Friedrich--Alexander Universit\"at Erlangen--N\"urnberg (FAU), DE-91058 Erlangen, Germany}


\date{\today}

\begin{abstract}
This article analyses the simulation methodology for wall-modeled large-eddy simulations using solvers based on the spectral-element method (SEM). To that end, algebraic wall modeling is implemented in the popular SEM solver Nek5000. It is combined with explicit subgrid-scale (SGS) modeling, which is shown to perform better than the high-frequency filtering traditionally used with the SEM. In particular, the Vreman model exhibits a good balance in terms stabilizing the simulations, yet retaining good resolution of the turbulent scales. Some difficulties associated with SEM simulations on relatively coarse grids are also revealed: jumps in derivatives across element boundaries, lack of convergence for weakly formulated boundary conditions, and the necessity for the SGS model as a damper for high-frequency modes. In spite of these, state-of-the-art accuracy is achieved for turbulent channel flow and flat-plate turbulent boundary layer flow cases, proving the SEM to be a an excellent numerical framework for massively-parallel high-order WMLES.
\end{abstract}

\maketitle




\section{Introduction} \label{sec:intro}

Wall-modeled LES (WMLES) is a turbulence modeling technique, in which the turbulent scales in the inner region of the turbulent boundary layer (TBL) are left unresolved by the computational grid and instead accounted for by a dedicated model.
So, if $\delta$ and $\delta_\nu$ are, respectively, the outer and inner length scales of the TBL, WMLES can be defined as resolving scales $\sim \delta$ as in a conventional wall-resolved LES (WRLES), and modeling the scales $\sim \delta_\nu$ separately.

This definition does not specify the concrete way of how the split between the resolved and modeled scales is realized through the choice of governing equations.
And indeed, several approaches are possible.
The interested reader is referred to~\cite{Larsson2016, Bose2018a} for a comprehensive review.
Following the classification introduced by Larsson et al.~\cite{Larsson2016}, the scope of this work is limited to wall-stress modeling, which is based on the assumption that providing the correct value of the wall shear stress $\tau_w$ is sufficient for accounting the effects of the inner layer dynamics on the rest of the flow.

The focus of this article is evaluating the predictive accuracy of WMLES conducted with a solver based on a spectral-element discretization.
The spectral element method (SEM), is an umbrella term for a high-order Galerkin finite element discretization scheme, typically with an intra-element node distribution that matches a Gaussian quadrature rule.
The latter is used to evaluate the integrals appearing in the weak form of the governing equations.
A major distinction can be made between SEM formulations based on continuous and discontinuous Galerkin (CG vs DG), which differ in the choice of the approximation space.
In the continuous case, the solution fields do not experience jumps across the element boundaries.
The particular SEM approach used here is implemented in the incompressible flow solver Nek5000 and is of the continuous type.
Additional information regarding numerical methods is provided in Section~\ref{sec:numerics} and for a comprehensive coverage the reader is referred to~\cite{Deville2002}.

The interest in using an SEM-based code for WMLES is chiefly motivated by two factors.
One is the exceptional parallel performance of such codes due to most operations being performed locally on a single element.
This applies to both traditional CPU-based computing, as well as GPU-accelerated~\cite{Offermans2016, Jansson2024}.
The other is the high order of the discretization, commonly chosen to be between 5\textsuperscript{th} and 9\textsuperscript{th}.
This potentially leads to better accuracy for the same number of degrees of freedom, although considering the fact that part of the TBL is not fully resolved in a WMLES, this is not guaranteed.

A brief overview of previous works on WMLES with the CG SEM is now given to put the current work into context.
Chatterjee and Peet~\cite{Chatterjee2017} used Nek5000 for simulating a neutral atmospheric boundary layer, focusing on the parametrisation of the Smagorinsky subgrid-scale model.
Pal et al.~\cite{Pal2021, Pal2022} presented Nek5000 results from channel flow simulations, including heat transfer.
Possibly the most important precursor to this work is the article of Gillyns et al.~\cite{Gillyns2022}, where the authors present channel flow results from Nek5000 simulations and study the effect of various WMLES parameters.
One should also mention the work of Huang et al.~\cite{Huang2022}, which is distinguished by the use of spectral vanishing viscosity and p-refinement.
Finally, the reader may also be interested in the following works on high-order WMLES, which treat the problem differently from us in terms of either the numerical method or choice of governing equations~\cite{Lodato2014, Frere2017, Frere2018, Wang2022}.

Building on these previous works, this study contributes the following to advance SEM-based WMLES further.
\begin{itemize}
	\item Besides for turbulent channel flow, we also show results from flat-plate zero-pressure-gradient TBL simulations.
	We show that the latter is significantly more difficult to compute accurately and highlight important issues regarding meshing developing boundary layers with structured meshes.
	\item In our work we use state-of-the-art algebraic subgrid scale models, Vreman~\cite{Vreman2004} and Sigma~\cite{Nicoud2011}.
	We demonstrate that these models can be used successfully in tandem with the CG SEM, at the same time highlighting certain issues stemming from the properties of the numerical method.
	\item Besides the traditional approach of using the Neumann boundary condition for wall-parallel velocity to prescribe the wall stress, we also test a subgrid viscosity-based approach (details in Section~\ref{sec:nut_condition}), previously not tested with the SEM.
	\item We provide an open-source implementation of the WMLES code at\\ \texttt{https://github.com/timofeymukha/nekwmles}.
\end{itemize}

The rest of the paper is structured as follows.
Section~\ref{sec:methods} provides a complete description of the WMLES methodology used to conduct the simulations.
Section~\ref{sec:channel} presents and discusses results from simulations of turbulent channel flow at $\rey_\tau \approx 8000$.
The same is done for flat-plate TBL simulations in Section~\ref{sec:tbl}.
Concluding remarks are given in Section~\ref{sec:conclusions}.

\section{Computational fluid dynamics methods} \label{sec:methods}
Here, we first present the LES governing equations (Section~\ref{sec:eqations}) and an overview of the numerical solution procedure (Section~\ref{sec:numerics}).
This is followed by a discussion of various aspects of the wall modeling methodology in Section~\ref{sec:wall-bc} and subgrid scale modeling in Section~\ref{sec:sgs}.

\subsection{Governing equations} \label{sec:eqations}
The governing equations for LES of incompressible flow are
\begin{align}
&  \pdiff{u_i}{t}+ u_j \pdiff{u_i}{x_j} = -\frac{1}{\rho} \pdiff{p}{x_i} + \pdiff{\tau_{ij}}{x_j} + f_{hp},\quad (i=1,2,3) \label{eq:les} \\
&  \pdiff{u_j}{x_j} = 0.
\end{align}
Here, $u_i$ and $p$ are the velocity and pressure fields implicitly filtered by the computational grid.
The tensor $\tau_{ij}$ combines the contributions of the viscous and subgrid-scale (SGS) stresses.
The latter are here modeled using the Boussinesq approximation, leading to
\begin{equation}
    \label{eq:viscstress}
    \tau_{ij} =(\nu + \nu_\mathrm{sgs})S_{ij}, \quad S_{ij} = \pdiff{u_i}{x_j} + \pdiff{u_j}{x_i}.
\end{equation}
Further discussion of SGS modeling is deferred to Section~\ref{sec:sgs}.

Finally, $f_{hp}$ on the right-hand-side of~\eqref{eq:les} is a filter-based relaxation term, which attenuates high-frequency modes in the solution field based on a modal expansion of the solution in each element:
\begin{equation} \label{eq:filter}
	f_{hp} = \chi(u - \tilde{u}),
\end{equation}
where $\chi$ is a selectable constant that controls the strength of the forcing and $\tilde{u}$ is a low-pass filtered velocity signal, which is parametrized by the number of modes affected by the filtering.
The filtering is generally necessary for the stability of the SEM, however, in the case of LES additional dissipation is provided by the SGS model.
Therefore, by default in our simulations the filtering is not applied.
However, following Gillyns et al.~\cite{Gillyns2022} we test using $f_{hp}$ as an implicit SGS model, see Section~\ref{sec:sgs} for details.

\subsection{Discretization and solution procedure} \label{sec:numerics}
As already mentioned in the introduction, the CG SEM is a high-order finite element method.
The computational domain is discretized into hexahedral elements, inside which the solution is represented as polynomials of the user-selected order, $N$.
Here, we use $N=7$, which is a fairly standard choice for high-fidelity CFD.
During the simulation, each element is mapped to a reference element of size $[-1, 1]^3$ on which the operations are performed before mapping the solution back to the original geometry.

In Nek5000, the solution polynomials are Lagrange interpolants, built using the values of the solution fields, stored in Gauss-Legendre-Lobatto (GLL) nodes.
Crucial for computational efficiency, it can be shown that the three-dimensional basis can be build as a tensor product of one-dimensional polynomials.
When necessary, the solution can also be represented in a hierarchical modal basis using Legendre polynomials.
For example, this is used in the relaxation filtering discussed above.

The advantage of using GLL nodes is that they also serve as the nodes of the Gaussian quadrature used to compute the integrals arising in the weak form of the governing equations.
This leads to a diagonal, and thus easily invertible, mass matrix.
The integration of the convective term requires special treatment to avoid instabilities.
In Nek5000, the 3/2 over-integration rule is used to that end~\cite{Malm2013}.

Since the velocity field is continuous across elements, global mass conservation is fulfilled when the velocity field is divergence-free locally in each element.
On the other hand, global momentum conservation \textit{is not guaranteed}, since the velocity derivatives may have jumps across element boundaries.
These jumps diminish with mesh refinement and become close to negligible at resolutions typical of wall-resolved simulations.
However, for WMLES the lack of momentum balance can be significant enough to deteriorate the accuracy of the results, as we show below.

A mixed scheme is used for time integration, applying third-order backward differencing to the viscous term and explicit extrapolation of the same order to the convective term~\cite{Karniadakis1991}.

The so-called $P_N$-$P_{N-2}$ formulation is adopted to solve the governing equations, which implies that the solution polynomials for pressure are two orders lower than that for velocity.
This is necessary to avoid spurious modes in the solution.
The discretized system of equations is solved using operator splitting by means of block LU decomposition.
The momentum equation is solved using a preconditioned conjugate gradient solver, whereas a generalized minimal residual method solver is used for the pressure-Poisson equation.
The additive overlapping Schwarz preconditioner is applied to the latter~\cite{Fischer1997}, and the XXT direct solver used as the coarse-grid solver.
Here XXT refers to the Cholesky decomposition of some matrix $A$ into $XX^T$, for details consult~\cite{Tufo2001}.

\subsection{Wall boundary conditions}  \label{sec:wall-bc}
\subsubsection{Wall modeling approach} \label{sec:wall-modeling}
The standard wall-stress modeling approach is considered here, in which the wall model predicts the value of the wall shear stress $\tau_w$ based on the LES solution in the outer layer of the TBL.
This value is then applied at the wall via an appropriate boundary condition.

Since we only study attached boundary layers in this work, Spalding's law of the wall~\cite{Spalding1961} can generally be expected to provide an accurate relationship between velocity and $\tau_w$.
Therefore, it is here used as the wall model.
Similar expressions, like Reichardt's law~\cite{Reichardt1951}, or simply the log-law could be used as well.
An analysis of the accuracy of Spalding's law is given in Appendix~\ref{sec:appendix_spalding}.

Collecting the input to the wall model is referred to as sampling.
An important parameter is the distance from the sampling point to the wall, denoted as $h$.
Since the wall models of the type used here are generally most accurate in the log-law region, $h$ should be selected such that $h^+ \gtrapprox 50$ and $h/\delta \lessapprox 0.3$.
At high Reynolds numbers, the former requirement is fulfilled even at the first off-wall grid point.
However, following~\cite{Kawai2012}, there is strong evidence that $h$ should be at least large enough to reach the second off-wall point, see e.g~\cite{Calafell2019, Mukha2019, Hosseinzade2021}.
We refer to~\cite{Kawai2012} for details, but the main reason for this is the necessarily poor prediction of the velocity value in the first off-wall node.
In the context of high-order methods, an important question is whether this implies that sampling from the whole near-wall \textit{element} is suboptimal.
The argument in favor of this being that the velocity values in the whole element are coupled to that in the first node.
However, in~\cite{Gillyns2022} sampling from the uppermost node in the near-wall element was used without issues.
In our implementation, arbitrary $h$ can be selected for each wall node.
The sampled value is obtained via spectral interpolation within the element where the sampling point happens to lie.

Another important choice is whether to average (in time or space) the sampled velocity signal prior to its input into the wall model.
The argument for doing so is the fact that the equations governing the wall model (e.g.~the log-law) are valid for the mean velocity and not the transient LES signal.
It is trivial to show that the fluctuations in the sampled velocity values, $u^{wm}$, will lead to an over-prediction of $\langle \tau_w \rangle$~\cite{Hultmark2013}.
Nevertheless, there are conflicting accounts in the literature regarding the real impact of averaging on predictive accuracy.
On one end of the spectrum, Yang et al.~\cite{Yang2017} demonstrated that averaging alone can remove the infamous log-layer mismatch (LLM), i.e.~the vertical shift in the inner-scaled mean velocity profile due to an erroneous $\tau_w$.
In the context of a non-equilibrium wall model, the issue is studied in~\cite{Calafell2019} and, more recently, in~\cite{Hosseinzade2021}.
Both works highlight the importance of averaging, although the positive effect is stronger when the sampling point is close to the wall.
This is logical, since the variance of the velocity signal decays with the wall-normal distance.
On the other hand,~\cite{Huang2022} reports no significant effect of time-averaging on the predictions.
In our implementation, we provide the option to apply exponential averaging, i.e.
\begin{align}
\label{eq:avrg1}
&    u_{n}^{wm} = \epsilon u_{n}^{wm} + (1- \epsilon)u_{n-1}^{wm},\\
\label{eq:avrg2}
& \epsilon = \Delta t / (\Delta t + \Delta t_\mathrm{avrg})
\end{align}
where $\Delta t$ is the time step and $\Delta t_\mathrm{avrg}$ is the averaging length.
Interestingly, in~\cite{Gillyns2022} the authors report that their Nek5000-based WMLES diverged when no time-averaging was applied.
We have not encountered this issue in our simulations.
Finally, in~\cite{Chatterjee2017}, explicit spatial filtering of high frequency modes from the velocity signal prior to input to the wall model is used. (Similar in effect to relaxation-based filtering discussed in Section~\ref{sec:eqations}.)
Here, we don't consider this possibility, but it could be easily incorporated because the filtering functionality is already a part of Nek5000.

\subsubsection{Weak form of the viscous stress term} \label{sec:weak_form_stress}
In the previous section, it was established that the wall modeling operates by predicting the correct wall shear stress.
This prediction should be realized at the wall via an appropriate boundary condition.
To that end, it is important to see how $\tau_w$ enters the weak formulation of the governing equations, from which SEM is derived.
Since the only relevant term in this context is that of the viscous stress tensor, we will consider it here in isolation, i.e.

\begin{equation}
    \frac{\partial \tau_{ij}}{\partial x_j} = 0, \quad i=1,2,3.
\end{equation}
The corresponding weak form is
\begin{equation}
	\int_\Omega  v_i \frac{\partial \tau_{ij}}{\partial x_j} \, d\xi = 0,
\end{equation}
where $v_i$ are test functions selected from the same approximation space as the solution, and $\Omega$ is the computational domain.
Now, using the following identity,
\begin{equation}
	\frac{\partial v_i \tau_{ij}}{\partial x_j} = \frac{\partial v_i}{\partial x_j} \tau_{ij} + v_i \frac{\partial \tau_{ij}}{\partial x_j},
\end{equation}
we obtain
\begin{equation}
	\int_\Omega   \frac{\partial v_i}{\partial x_j} \tau_{ij} \, d\xi = \int_\Omega  \frac{\partial v_i \tau_{ij}}{\partial x_j}\, d\xi.
\end{equation}
Applying the Gauss-Ostrogradsky theorem to the right-hand side we get
\begin{equation}
	\int_\Omega  \frac{\partial v_i}{\partial x_j} \tau_{ij}\, d\xi = \int_{\partial \Omega}   v_i \tau_{ij} n_j\, d\xi.
\end{equation}
Now let $\tau_i = \tau_{ij}n_j$ and thus $\tau_{w, i} = \tau_i - (\tau_j n_j)n_i$, leading to
\begin{equation}
	\label{eq:stress}
	\int_\Omega  \frac{\partial v_i}{\partial x_j} \tau_{ij}\, d\xi = \int_{\partial \Omega} \left[  v_i \tau_{w, i} +  (\tau_j n_j)v_i n_i \right] \, d\xi.
\end{equation}
The second term under the integral on the right-hand side is equal to zero because of the strongly enforced non-penetration condition, $v_in_i = 0$.
The first term depends on the boundary condition.
In case of the standard no-slip condition, it vanishes since the test functions are zero-valued at the wall.
In the case of a Neumann condition, it is non-zero.

\subsubsection{Neumann boundary condition} \label{sec:neumann_condition}
The above derivation demonstrated that the predictions of the wall model can be incorporated into the momentum balance via a source term that arises naturally due to the integration by parts procedure.
Since the viscosity at the wall is set, this is essentially a Neumann-type condition for the wall-parallel velocity components.
In combination with a homogeneous Dirichlet condition for wall-normal velocity, this is, perhaps, the most natural way of prescribing the correct stress at the boundary.

This boundary condition offers the possibility to set the direction of $\tau_w$ by choosing the magnitude of its two components.
The standard choice is to align $\tau_w$ with the sampled wall-parallel velocity, which is also what we implemented in our code.

%

As derived in the previous subsection, the Neumann condition is enforced weakly.
This means that  the local wall shear does not necessarily equal to what is predicted by the wall model.
This equality is only achieved asymptotically as the mesh is refined.
Since in WMLES the inner region is, by definition, unresolved, one generally ends up with a discrepancy.
We will show that its magnitude can be very large on coarser WMLES grids and that it depends heavily on the subgrid scale modeling.
Note that this issue is unique to the finite element method, and can come as a surprise to those transitioning to an SEM solver from e.g.~a finite volume-based code.

As a consequence of the above, it is possible to report two values of $\tau_w$, one predicted by the wall model, $\tau^{wm}_w$, and the other one actually observed at the wall, $\tau^{wall}_w$.
This is usually not discussed, but the standard approach appears to be reporting the value from the wall model.
The important question is which value is important dynamically, and it will be demonstrated that it is, in fact, $\tau^{wm}_w$. 
For example, in channel flow $\tau^{wm}_w$ correctly scales the Reynolds shear stress to produce the right slope of its linear profile.
Nevertheless, it could be considered unsatisfactory and somewhat confusing that the wall-normal derivative of the obtained mean velocity field cannot be used to compute the wall stress a posteriori.


\subsubsection{Viscosity-based boundary condition} \label{sec:nut_condition}
It would be beneficial to have an alternative boundary condition, which would be enforced strongly. 
Assuming here and below that $x$ and $z$ point, respectively, in the streamwise and spanwise direction, and $y$ in the wall-normal direction, the magnitude of the wall shear stress is expressed as
\begin{equation}
    \label{eq:tau_def}
     |\tau_w| = (\nu + \nu_\mathrm{sgs})\sqrt{S^2_{xy} + S^2_{zy}}.
\end{equation}
It follows, that given some $|\tau^{wm}_w|$ from the wall model, a way to set it at the wall is to manipulate the subgrid viscosity:
\begin{equation} \label{eq:tau_sgs}
	\nu_\mathrm{sgs} = |\tau^{wm}_w|/\sqrt{S^2_{xy} + S^2_{zy}} - \nu
\end{equation}
This is combined with setting a no-slip condition for the velocity.
Note that since all the boundary conditions are now Dirichlet, the stress is enforced strongly.

Another advantage of this boundary condition is that it allows to easily combine wall-resolved and wall-modeled LES.
If the grid is sufficiently refined for resolving the inner layer, $\nu_\mathrm{sgs}$ can simply be set to zero at the wall, resulting in an ordinary no-slip condition for velocity.
Combining WRLES and WMLES in this manner is performed in~\cite{DeVanna2021}, in the context of compressible flow simulations using finite differences.

This boundary condition has long been used in the finite volume code OpenFOAM, the reason mainly being a better fit into the code structure of the software rather than mathematical considerations.
For WMLES results, the reader is referred to~\cite{Mukha2019, Rezaeiravesh2019a, Mukha2021}.
A study comparing the Neumann and the viscosity-based condition can be found in~\cite{Bae2021a}, with a finite difference code used to perform the simulations.
The viscosity-based condition is found to provide somewhat more accurate results.
However, to our knowledge, the viscosity-based approach has not been tried in the framework of the SEM.

Note that since $\nu_\mathrm{sgs}$ is a scalar, the direction of $\tau_w$ cannot be controlled, and is determined by the local values of $S_{xy}$ and $S_{zy}$.
It is illustrative to consider channel flow, where $\mean{\tau_w}$ is aligned with the mean flow direction, $x$.
Assume that the sampled velocity vector changes direction, but happens to always coincide in magnitude with the correct mean value of velocity for the selected $h$.
Then, given the validity of the model's governing equation (e.g.~the log-law), it will at each time-step predict $|\tau^{wm}_w| = |\mean{\tau_w}|$.
However, when setting the direction of $\tau^{wm}_w$, the component $\tau^{wm}_{w,z}$ will be non-zero, due to a non-zero $S_{zy}$.
Consequently, $\mean{\tau^{wm}_{w,x}}$ will be less than $|\mean{\tau_w}|$ in spite of the wall model giving a perfect prediction of the magnitude of the mean stress at each time step.

\subsection{Subgrid scale modeling} \label{sec:sgs}
Subgrid scale modeling is acknowledged to be difficult in WMLES due to the poor performance of existing models on coarse grids near the wall.
Generally, the assumption behind the models is that they are to compensate for the near-isotropic turbulent scales that are not resolved on the LES grid.
However, for WMLES this assumption breaks down because in the inner layer the even the resolved velocity fluctuations may be largely unphysical.
And since the error structure there strongly depends on the wall model and the numerical methodology, it is not easy to universally formulate what the objective of SGS modeling is exactly.

There is also a tendency to not use explicit SGS models at all in SEM-based codes.
For WMLES, Gillyns et al.~\cite{Gillyns2022} and Pal et al.~\cite{Pal2022} used filtering of high-frequency modes, whereas Huang et al.~\cite{Huang2022} used spectral vanishing viscosity.
No explicit model is used in the discontinuous Galerkin-based simulations in~\cite{Frere2017, Frere2018}.
However, the use of a mixed scale-similarity model together with the SEM in~\cite{Lodato2014} should be mentioned.
Also, a wall-damped Smagorinsky model has been used in the Nek5000 simulations in~\cite{Chatterjee2017}, which is perhaps partially explained by the fact that the authors are interested in atmospheric boundary layers.
(The same authors used filtering instead in a later work~\cite{Chatterjee2018}, but for WRLES.)

The meteorological community are some of the most active users of WMLES, since one often considers the limit of an infinite Reynolds number.
Explicit SGS modeling is common, and there are several works exploring the effects of various models on the predictive accuracy, see, for example,~\cite{Tkachenko2021, Gadde2021}.
It is not uncommon to use quite advanced and computationally complex models, like the LASD~\cite{Bou-Zeid2005}.

On the other hand, the engineering community seems to have gravitated towards simple algebraic models, like the Vreman model~\cite{Vreman2004}, WALE~\cite{Nicoud1999a}, and its successor, the Sigma model~\cite{Nicoud2011}.
For WMLES, the Vreman model is used in~\cite{Owen2019, Goc2020, Goc2021}, and WALE in, for example,~\cite{Mukha2019, Blanchard2021}.

In this study, we will use the Vreman and Sigma models in order to evaluate the performance of explicit SGS modeling in SEM-based WMLES.
Additionally, we will follow Gillyns et al.~\cite{Gillyns2022} and use the high-pass filter~\eqref{eq:filter} to perform implicit WMLES.
We note that in our implementation the value of the SGS viscosity is not forced to zero at the wall.
The desire to compare Vreman and Sigma chiefly stems from the different behavior of those models in the near-wall region.
In the Sigma model (and also WALE), $\nu_\mathrm{sgs}$ asymptotically behaves as a cubic function of the wall-normal distance.
On the other hand, in the Vreman model, the asymptotic behavior is linear~\cite{Nicoud2011}.

The importance of the near-wall behavior of the subgrid model is highlighted in the work of Brasseur and Wei~\cite{Brasseur2010}.
This article offers, perhaps, the most rigorous analysis of the causes of the log-layer mismatch and defines modeling criteria that have to be fulfilled to avoid it.
It is in this framework that Chatterjee and Peet calibrated the wall-damping of the Smagorinsky model in their Nek5000 simulations of the atmospheric boundary layer~\cite{Chatterjee2017}.
Brasseur and Wei show that any SGS model introduces a spurious length scale, $\delta_\mathrm{sgs}$, which manifests itself as an ``LES buffer layer'' in the velocity profile, leading to an overshoot in the near-wall velocity gradient.
According to the authors, this effect can be suppressed by ensuring that the resolved shear stress dominates its SGS counterpart in the first off-wall grid node.
Naturally, the value of this ratio is dependent on the strength of the SGS model near the wall.

The arguments of Brasseur and Wei~\cite{Brasseur2010} in favor of smaller values of $\nu_\mathrm{sgs}$ near the wall have, however, been disputed.
In~\cite{Jaegle2010}, the authors observe that, when a Neumann condition is used at the wall, there is a direct relationship between the slip velocity and the value of $\nu_\mathrm{sgs}$ at the wall.
Lower viscosity leads to lower slip velocities and, potentially, even negative velocities at the wall.
This can lead to instabilities.
In~\cite{Blanchard2021}, the authors show that WALE and Sigma perform very poorly in channel flow WMLES, leading to strong over-prediction of the wall stress.
(The paper is dedicated to a remedy for this.)
We note, however, that in~\cite{Mukha2019,Rezaeiravesh2019a, Mukha2021} the WALE model performed well for WMLES.
It must be concluded that the accuracy is a strong function of the combination of the boundary conditions, SGS model and numerical method.

Finally, we briefly discuss the procedure to compute the SGS length scale.
On isotropic grids, the most straightforward choice for SEM is to use the local spacing between GLL points.
This roughly corresponds to the cubic root of the cell volume in finite volume codes.
However, recall that the GLL distribution of nodes introduces a resolution bias towards the edges of the element.
In this work, we take a somewhat conservative approach and assign the maximum length scale computed based on GLL spacing to the entire element.

\section{Fully developed turbulent channel flow simulations } \label{sec:channel}

\subsection{Case setup}
We consider fully developed turbulent channel flow at $\rey_\tau \approx 8000$.
DNS data~\cite{Yamamoto2018} are used as reference.
The size of the computational domain is $8\delta \times 2\delta \times 6 \delta$.
The flow is forced to a fixed volumetric flow rate, defined by the bulk velocity, $U_b$.
The default settings for the wall model are: $h=0.2\delta$, no time averaging for $u^{wm}$, $\kappa=0.387$, $B=4.2$.
The Spalding's law constants are based on what is reported for the DNS data in~\cite{Yamamoto2018}.
All simulations are initialized by mapping velocity fields from previous channel flow simulations and run at Courant number less than 0.5.
Statistical quantities are obtained by averaging over 20$\delta/u_\tau$.

The simulations are performed on a set of 3 meshes.
The first mesh M1 corresponds to node resolution $\approx \delta/15$ in all directions, although some modulation of this value is introduced by the GLL distribution of the nodes within each element.
The mesh M2 is produced from M1 by doubling the number of elements in all three directions, and for M3 the same is done again.
Meshes M2 and M3 can be considered fine by engineering WMLES standards.
More information about the meshes can be found in Table~\ref{tab:channel_mesh}.

\begin{table}[htp!]
	\caption{Computational meshes for channel flow simulations. $N_x$, $N_y$, and $N_z$ are the number of elements along the three coordinate axes. $N_{gll}$ is the total number of nodes in the mesh.}
	\label{tab:channel_mesh}
        \begin{ruledtabular}
	\begin{tabular}{p{1cm}p{1cm}p{1cm}p{1cm}p{2cm}}
		Mesh & $N_x$ & $N_y$ & $N_z$ & $N_{gll}$ \\
		\noalign{\smallskip}\hline\noalign{\smallskip}
		M1 & 16 & 4 & 12 & $0.39 \cdot 10^6$ \\
		M2 & 32 & 8 & 24 & $3.14 \cdot 10^6$ \\
		M3 & 64  & 16 & 48 & $25.17 \cdot 10^6$ \\
	\end{tabular}
        \end{ruledtabular}
\end{table}

\subsection{Simulations using the Neumann boundary condition} \label{sec:channel_neumann}

\subsubsection{Velocity statistics in outer scaling}
We begin with the results obtained using the Neumann boundary condition and the Vreman subgrid model.
The main purpose of the analysis here is to demonstrate the overall level of accuracy of the velocity statistics and their dependence on grid resolution.

\begin{figure}[htp!]
	\centering
	\includegraphics{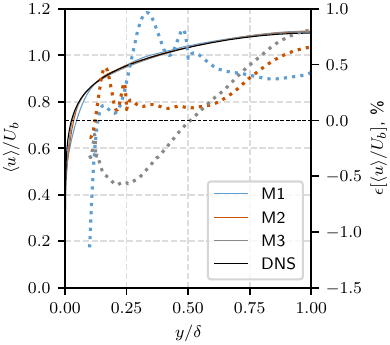}
	\caption{\textit{Solid lines:} Mean velocity profiles obtained in channel flow simulations, using the Neumann boundary condition. \textit{Dotted lines:} The relative errors in the profiles with respect to the reference DNS, in the outer layer.}
	\label{fig:channel_neumann_u}
\end{figure}

Figure~\ref{fig:channel_neumann_u} shows the mean velocity profiles as well as the relative errors in these profiles with respect to the DNS data (note the extra ordinate axis on the right of the plot).
Very good accuracy is obtained for $y>0.1\delta$ for all three grids, with the relative error not exceeding $1\%$.
This is certainly a positive outcome, but not an unexpected one, as similar accuracy has been obtained with Nek5000 by Gillyns et al.~\cite{Gillyns2022}.
Finite volume-based simulations also demonstrate a similar level of agreement, both on structured~\cite{Rezaeiravesh2019a} and unstructured~\cite{Mukha2021} grids.
However, the results here are more robust with respect to grid resolution.

Recall that on the M1 grid $\delta$ is meshed using only 2 elements.
The entire core of the channel is thus reasonably well represented on a single element.
There is no gain in accuracy from refining M2 to M3, and, in fact, the M2 profile exhibits the smallest error overall.
Close to the wall, we see that the M1 profile under-predicts the DNS appreciably, so capturing the velocity gradient in this region requires better resolution.

Interestingly, Gillyns et al.~\cite{Gillyns2022} used different resolutions in their simulations, depending on the Re number of the channel flow.
For $\rey_\tau = 5200$, the resolution was increased, testing 10 and 20 elements per $\delta$, which is denser than the M3 grid here.
The motivation for this is not entirely clear, but could possibly be connected to the fact that they don't use an explicit SGS model and thus lack the necessary dissipation to keep the simulation stable.
We return to this discussion later on.

As a final note on the mean velocity profiles, it is observed that at the sampling point, $h=0.2\delta$, the M1 and M2 profiles happen to exhibit virtually no error, whereas a small under-prediction is seen in the M3 curve.
We can thus generally expect the $\mean{\tau^{wm}_w}$ predictions to be accurate, as will be demonstrated below.

\begin{figure}[htp!]
	\centering
	\includegraphics{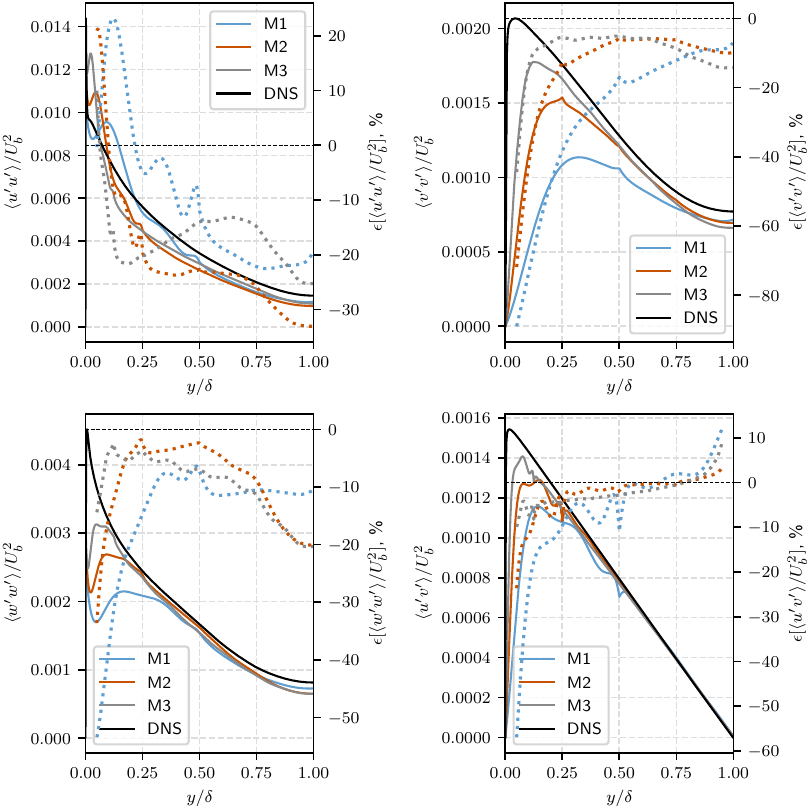}
	\caption{\textit{Solid lines:} Reynolds stress profiles obtained in channel flow simulations, using the Neumann boundary condition. \textit{Dotted lines:} The relative errors in the profiles with respect to the reference DNS.}
	\label{fig:channel_neumann_re}
\end{figure}

The attention of the reader is now directed towards Figure~\ref{fig:channel_neumann_re}, which shows the obtained Reynolds stresses together with the corresponding error profiles.
The first immediate observation is that in the core of the channel the errors in the normal stresses are an order of magnitude higher than that for velocity.
This is a result typical of WMLES, as is the pattern of near-wall over-prediction of $\mean{u'u'}$ and under-prediction of $\mean{v'v'}$ and $\mean{w'w'}$.
The strength of the latter is what distinguishes the results on different grids most.
In contrast to what has been observed for $\mean{u}$, there is a strong jump in accuracy from M2 to M3.
The M1 profile lies significantly below the DNS, for $\mean{v'v'}$ all the way to $y/\delta \approx 0.5$.
On the other hand, away from the wall, M2 and M3 profiles agree very well, and the reduction in accuracy for M1 is not very strong.

The discussion above is generally applicable for the shear stress as well, with the modification that in the core of the channel the agreement is excellent.
This is largely an effect of the fact that the linear profile of $\mean{u'v'}$ is predetermined by the flow configuration ($\mean{u'v'} \approx |\mean{\tau_w}|(1 - y/\delta)$ in the core of the channel).
In other words, the accurate results indicate that the wall shear stress is predicted correctly.
We proceed to examine these predictions sin the next subsection.

\subsubsection{Wall shear stress predictions and the log-layer mismatch}

We now examine the accuracy of the wall modeling by looking at the values of the wall shear stress.
There is a set of 3 values that we can consider for each simulation: The prediction of the wall model, $\mean{\tau_w^{wm}}$; The value at the boundary computed using the gradient of the mean velocity profile, $\mean{\tau_w^{wall}}$; The value derived from the mean pressure gradient driving the flow, $\mean{\tau_w^{\nabla p}}$.
These are reported in Table~\ref{tab:neumann_tau}.

\begin{table}[htp!]
	\caption{Relative errors in the values of the wall shear stress obtained in channel flow simulation using the Neumann boundary condition.}
	\label{tab:neumann_tau}       
	\begin{ruledtabular}

	\begin{tabular}{p{2cm}p{2cm}p{2cm}p{2cm}}
		Mesh & $\epsilon[\mean{\tau_w^{wm}}]$ & $\epsilon[\mean{\tau_w^{wall}}]$ &$\epsilon[\mean{\tau_w^{\nabla p}}]$ \\
		\noalign{\smallskip}\hline\noalign{\smallskip}
		M1 & 0.71\% & 12.58\% &  1.95\% \\
		M2 & 0.76\% & 10.86\% &  0.96\% \\
		M3 & -0.54\% & 8.07\% & -3.84\% \\
	\end{tabular}
	\end{ruledtabular}
\end{table}

\begin{figure}[htp!]
	\centering
	\includegraphics{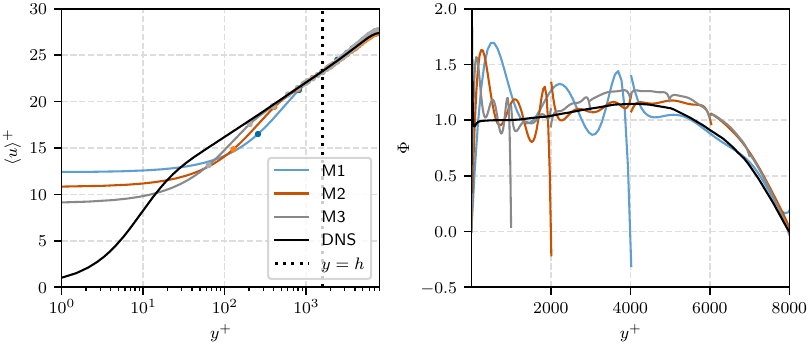}
	\caption{\textit{Left:} Inner-scaled mean velocity profiles obtained in channel flow simulations, using the Neumann boundary condition. \textit{Right:} The profile of the indicator function $\Phi$ from the same simulations.}
	\label{fig:channel_neumann_u_inner}
\end{figure}

We see that the predictions of the wall model are very accurate on all meshes.
This is reflected in the left plot in Figure~\ref{fig:channel_neumann_u_inner}, where the velocity profiles are shown in inner scaling.
In line with the theory in Brasseur and Wei~\cite{Brasseur2010}, an ``LES buffer layer'' can be clearly seen, as well as the reduction of its size with mesh refinement.
On the other hand, no LLM is observed on any of the meshes, even though the shear stress is clearly smaller than the subgrid one near the wall on, for example, mesh M1 (see Figure~\ref{fig:channel_neumann_re}).
It therefore appears that the dominance of the resolved shear stress over the modeled one near the wall is not a necessary condition for capturing the log-layer.

An alternative way of analyzing how well the simulations capture the law of the wall is by looking at the indicator function $\Phi = \kappa y^+ \text{d}\mean{u}^+/\text{d}y^+$.
The profiles of this quantity are shown in the right plot of Figure~\ref{fig:channel_neumann_u_inner}.
They clearly expose the fact that the derivatives of velocity are generally discontinuous across elements.
In SEM, diminished jumps in the derivatives can be considered as an indicator of mesh convergence.
From this point of view, convergence is achieved for $y > 0.25$ on M2 and M3.
In fact, these two profiles overlap well in a region closer to the center of the channel.
However, $y \approx 0.25$ is more or less the edge of the log-layer, which implies that the layer itself is not resolved to the extent, which would, perhaps, be desirable.
The oscillatory behavior of the derivative close to the wall also makes it difficult to discern the peak in $\Phi$ associated with the LES buffer layer as opposed to purely numerical artifacts.
We continue the analysis of $\Phi$ when comparing results obtained with different SGS modeling strategies.

Returning to the data in Table~\ref{tab:neumann_tau}, we see that the stress values at the boundary, $\mean{\tau_w^{wall}}$, over-predict the reference data, and become increasingly more accurate with mesh refinement.
The latter should, in principle, be considered coincidental, since we are, by design, too far from approaching the WRLES limit on all three considered meshes.
Clearly, $\mean{\tau_w^{wall}}$ values are not indicative of the simulation accuracy and should essentially be disregarded.

Finally, $\mean{\tau_w^{\nabla p}}$, which is obtained from the magnitude of the force driving the flow, does not show perfect agreement with the prediction of the wall model, $\mean{\tau_w^{wm}}$.
This is interesting, because, as discussed in Section~\ref{sec:weak_form_stress}, $\tau_w^{wm}$ also enters the momentum equation as a source term, so, in principle, the driving pressure gradient should compensate that exactly.
The deviation from that is, therefore, a direct measure of global momentum imbalance in the discretized system and possible errors due to time-averaging.
As we see from the table, this error is of the same order of magnitude as that in the wall shear stress predictions made by the wall model.

\subsubsection{Time-averaging of the wall model input}
We now consider the effect of averaging the velocity signal used by the wall model, see Eqs.~\eqref{eq:avrg1}-\eqref{eq:avrg2}.
Before proceeding, it is helpful to recall the following properties of Spalding's law (see Appendix~\ref{sec:appendix_spalding}).
\begin{itemize}
	\item The law is not exact at $y=h$, which leads to an error in the wall stress prediction when the DNS $\mean{u}|_{y=h}$ is used as input.
	The error is only $-0.04$\% for $h=0.2\delta$.
	Nevertheless, Spalding's law will give a small accuracy boost to input velocities that over-predict the DNS due to error cancellation.

	\item The error in $\tau^{wm}_w$ is about twice the size of the error in $u^{wm}$. Of course, this is only a rule of thumb valid for errors around 1\%-2\%.

\end{itemize}

To start with, an a priori analysis  of the effects of averaging can be performed.
The first step involves using the mean velocity profiles generated by the simulations presented above.
In particular, we can use the values of $\mean{u}$ at $y=h$ as $u^{wm}$ to make a prediction of the wall stress.
This roughly corresponds to an infinite $\Delta t_\mathrm{avrg}$ in Eq.~\eqref{eq:avrg2}, under the assumption that the induced change in  $\tau^{wm}_w$ would not propagate back to a change in the velocity values.
This assumption generally holds well, see e.g.~the analysis in~\cite{Rezaeiravesh2019a}.
The wall stress computed this way is referred to as $\mean{\tau^{apri}_w}$.

Table~\ref{tab:apriori_tau} shows the values of the error in $\mean{\tau^{apri}_w}$, and, for convenience, of those obtained in the simulations, $\mean{\tau^{wm}_w}$ (the same as in Table~\ref{tab:neumann_tau}).
On meshes M1 and M2 the results are improved by using the mean velocity, but on M3 it becomes worse.
The reason for that is simple.
On all three meshes, using the transient signal shifts the error towards over-prediction (as expected, see Section~\ref{sec:wall-modeling}).
On mesh M3, $\epsilon[\mean{\tau^{apri}_w}] < 0$ so this shift leads to error cancellation.
The table also contains the value of $\epsilon[\mean{u}|_{y=h}]$, and for M2 and M3 it is about half the error in the stress, as expected.
For M1 the errors are so small that the rule of thumb does not hold.

\begin{table}[htp!]
	\caption{A priori analysis of the effects of time-averaging on the errors in $\mean{\tau_w}$. The columns show relative errors in: the mean wall stress computed by the wall model during the simulation, $\epsilon[\mean{\tau^{wm}_w}]$ (same as in Table~\ref{tab:neumann_tau}); the mean wall stress computed by the wall model using the mean velocity value, $\epsilon[\mean{\tau^{apri}_w}]$; the mean velocity value at $y=h$, $\epsilon[\mean{u}|_{y=h}]$.}
	\label{tab:apriori_tau}       
        \begin{ruledtabular}

	\begin{tabular}{p{2cm}p{2cm}p{2cm}p{2cm}}
		Mesh & $\epsilon[\mean{\tau^{wm}_w}]$ & $\epsilon[\mean{\tau^{apri}_w}]$ &$\epsilon[\mean{u}|_{y=h}]$ \\
		\noalign{\smallskip}\hline\noalign{\smallskip}
		M1 & 0.71\% & 0.06\% &  0.06\% \\
		M2 & 0.76\% & 0.2\% &  0.13\% \\
		M3 & -0.54\% & -1.07\% & -0.57\% \\
	\end{tabular}
        \end{ruledtabular}
\end{table}

While the above result demonstrates the benefits of averaging, it does not provide a guideline for selecting the averaging length, $\Delta t_\mathrm{avrg}$.
To analyze this, velocity values have been sampled at the sampling height during a simulation on mesh M2.
A probe was placed above each GLL wall node, and sampling was performed at every third time step.
The dataset thus faithfully reproduces the velocity data used by the wall model during the simulation.

\begin{figure}[htp!]
	\centering
	\includegraphics{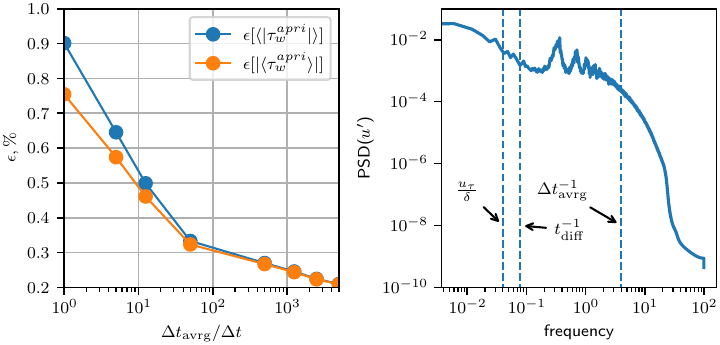}
	\caption{\textit{Left}: The dependency of $\epsilon[\mean{|\tau^{apri}_w|}]$ and $\epsilon[|\mean{\tau^{apri}_{w}}|]$ on $\Delta t_\mathrm{avrg}$.
	\textit{Right}: The temporal power spectral density of the streamwise velocity fluctuations at $y=h$.}
	\label{fig:channel_tau_fluct}
\end{figure}

Recall that the wall model first predicts the magnitude of the wall stress and then predicts the $x$ and $z$ components by making the stress vector parallel to that of the sampled velocity.
Using the dataset above, we can look at how both $\mean{|\tau^{apri}_w|}$ and $|\mean{\tau^{apri}_{w}}|$ depend on $\Delta t_\mathrm{avrg}$.
For the latter, the range from 0 to $5\cdot10^3 \Delta t$ is considered, where the upper bound is equal to $\delta/u_\tau$.
The results, in terms of the error with respect to the DNS, are shown in the left plot in Figure~\ref{fig:channel_tau_fluct}.
As expected, for $|\mean{\tau^{apri}_{w}}|$, the values go down from  $\approx 0.76$ to $\approx 0.2$, in line with the data in Table~\ref{tab:apriori_tau}.
Most of the gain from averaging is obtained by $\Delta t_\mathrm{avrg}/\Delta t = 50$, at which point the error reduction becomes slower.
It is also for this $\Delta t_\mathrm{avrg}$ that we observe the beginning of the overlap between $\mean{|\tau^{apri}_w|}$ and $|\mean{\tau^{apri}_{w}}|$.
This means that the $z$ component of velocity is essentially averaged-out at this point, and the direction of the stress is aligned with the mean flow.
It can therefore be conjectured that this is the main reason for the quick reduction of error, whereas subsequent, slower, reduction is attributed solely to eliminating the fluctuations in the streamwise velocity.
This agrees well with Yang et al.~\cite{Yang2017} that consider a forced correlation between the fluctuations of the stress and velocity vector to be the major cause of the log layer mismatch. 

The value $\Delta t_\mathrm{avrg} = 50\Delta t$ corresponds to $0.01 \delta/u_\tau$ in this study.
In~\cite{Gillyns2022}, the authors used a value 4 times lower.
Another study by Hosseinzade et al.~\cite{Hosseinzade2021} suggests relating the averaging length to the turbulent diffusion timescale $t_\mathrm{diff} = h/(\kappa u_\tau)$ and concludes that $\Delta t_\mathrm{avrg} = t_\mathrm{diff}$ is a robust choice.
Here, $t_\mathrm{diff} \approx 0.51 \delta/u_\tau$, so a significantly longer period than $50\Delta t$.

Using a time scale $\sim h$ is physically justified because the correlation between the wall stress and sampled velocity is stronger closer to the wall.
However, in WMLES the near-wall fluctuations of velocity are not predicted with sufficient accuracy to draw benefit from that, see Figure~\ref{fig:channel_neumann_re}.
Consequently, for smaller $h$ one may in practice need larger $\Delta t_\mathrm{avrg}$ to get good predictions of the stress.

In~\cite{Calafell2019}, Calafell et al.~instead suggest setting $\Delta t_\mathrm{avrg}$ to be large enough to filter out frequencies above the inertial subrange.
The right plot in Figure~\ref{fig:channel_tau_fluct} shows the temporal power spectral density of $u'$.
Both $u_\tau/\delta$ and $t_\mathrm{diff}$ correspond to large-scale motions, whereas $\Delta t_\mathrm{avrg} = 50\Delta t$ is located within the inertial subrange and therefore agrees well with the recommendation in~\cite{Calafell2019}.
For simple statistically steady-state cases like channel flow, a larger $\Delta t_\mathrm{avrg}$ can be selected, but generally it is important that the sampled velocity captures large-scale transients inherent to the flow.
A time scale based on the spectrum is difficult to select a priori, but our results indicate that the temporal stability of the direction of $\tau_w^{wm}$ may be used as a proxy criterion.
The latter, could in principle be evaluated during the simulation, opening up for the possibility of selecting $\Delta t_\mathrm{avrg}$ automatically.

\begin{figure}[htp!]
	\centering
	\includegraphics{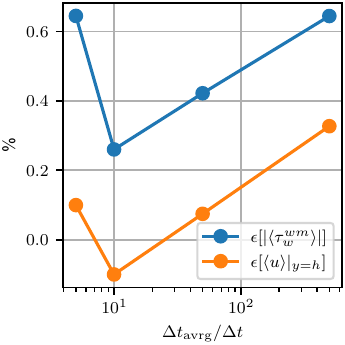}
	\caption{Errors in the mean wall shear stress and streamwise velocity as a function of the time averaging length of the sampled velocity signal.}
	\label{fig:channel_tau_fluct_post}
\end{figure}

To complement the a priori analysis, four WMLES with $\Delta t_\mathrm{avrg}$ set to $5\Delta t$, $10\Delta t$, $50\Delta t$, and $500\Delta t$ have been performed. The resulting errors in the wall stress are shown in Figure~\ref{fig:channel_tau_fluct_post}.
Unfortunately, no systematic accuracy gain is observed when increasing $\Delta t_\mathrm{avrg}$.
The error in the mean velocity at the sampling point changes with $\Delta t_\mathrm{avrg}$, violating the assumption of the a priori study.
The effect is generally small, but so are the errors in the stress.
At $\Delta t_\mathrm{avrg} = 500\Delta t$ we see that the usual rule of thumb of the error in the stress being twice of that in velocity holds.
At low $\Delta t_\mathrm{avrg}$, the velocity fluctuations push the error in the stress upwards.

The results of the a posteriori study do not imply that time-averaging of the velocity signal should not be used.
It is still the methodologically correct thing to do, and helps correctly isolate the source of the error in the predictions of the wall shear stress.
However, for algebraic models, a significant accuracy gain can generally not be expected, and the main factor determining the accuracy of the wall model is the accuracy of $u|_{y=h}$.
In terms of the wall model parameters, this means that $h$ will typically be more important than $\Delta t_\mathrm{avrg}$.
This conclusion is at odds with that of Yang et al.~\cite{Yang2017}, who claim that time averaging is critical for accurate stress predictions.
On the other hand, we acknowledge that for RANS-based wall models, which were studied by Calafell et al.~\cite{Calafell2019} and Hosseinzade et al.~\cite{Hosseinzade2021}, time averaging may indeed play a more prominent role.

\subsection{Comparing subgrid scale modeling approaches}
In this section, we compare the results obtained using the three considered SGS modeling approaches.
However, first the choices of the selectable constants for all models are given.
For the Vreman model, a single model constant should be selected, which is computed as $2.5 C_s^2$, where $C_s$ is the constant in the standard Smagorinsky model~\cite{Vreman2004}.
We use $C_s = 0.16$, which is slightly below the theoretical value for homogeneous isotropic turbulence.
The Sigma model is also controlled by one constant, which we set to $C_\sigma = 1.35$, following the recommendation in~\cite{Nicoud2011}.
Finally, the high-pass filter, see Eq.~\eqref{eq:filter}, is controlled by two parameters: $\chi$ and the number of modes in the low-pass filter.
Gillyns et al.~\cite{Gillyns2022} use $\chi=10$ and filter only the highest mode.
We find these settings to be more suitable for DNS, where the filter is used purely for numerical stability.
In fact,~\cite{Gillyns2022} reports that their solver diverged when no time-averaging was employed in the wall modeling procedure.
This indicates the need for stronger filtering.
Here, we filter two modes and set $\chi=100$.

\begin{figure}[htp!]
	\centering
	\includegraphics{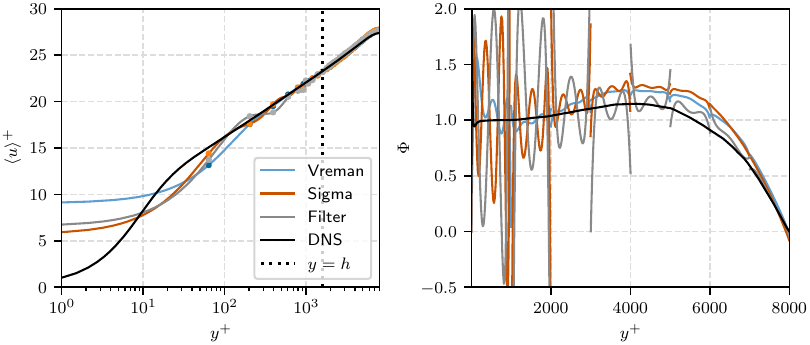}
	\caption{\textit{Left:} Inner-scaled mean velocity profiles obtained in channel flow simulations with different SGS models on the M3 grid. \textit{Right:} The profiles of the indicator function $\Phi$ from the same simulations.}
	\label{fig:channel_sgs_u_inner}
\end{figure}

The choice of the model has very little effect on the velocity profiles in the core of the channel, which is why we focus the discussion on the near-wall region (see Figure~\ref{fig:channel_sgs_u_inner}).
For brevity, we only consider the results on the densest mesh M3.
The first observation is that $\mean{\tau^{wm}_w}$ is accurate irrespective of the SGS model, and no log-layer mismatch is observed.
Additionally, lower dissipation near the wall leads to a more oscillatory solution, yet a smaller~$\delta_\mathrm{sgs}$.

The fluctuations in $\mean{u}$ propagate to its derivative, as indicated by the plot of the indicator function $\Phi$ in the right plot of Figure~\ref{fig:channel_sgs_u_inner}.
The behavior of $\Phi$ is clearly dominated by numerical errors in the inner layer.
It is impossible to analyze the overshoot in $\Phi$ in terms of the LES buffer layer.
In the outer layer, the jumps in the derivative also increase with decreased dissipation, remaining quite large in the case of the high-pass filtering.

\begin{figure}[htp!]
	\centering
	\includegraphics{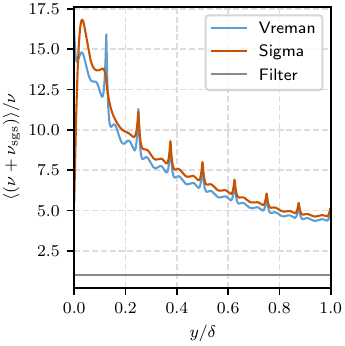}
	\caption{The total viscosity profiles in channel flow simulations using different SGS models.}
	\label{fig:channel_sgs_nu}
\end{figure}

The jumps and spikes in the derivatives have a direct impact on the SGS model, since the latter depends on the strain rate tensor.
Figure~\ref{fig:channel_sgs_nu} shows the profiles of the total viscosity in each of the three simulations.
As can be expected, spikes are present in these profiles as well.
Clearly, there is no correlation here with the resolution of turbulent structures; these peaks are just numerical artifacts.
Ensuring continuity of velocity derivatives can therefore be seen as a missing component in the current WMLES methodology and is a subject of future work.

One clearly sees the decrease in $\nu_\mathrm{sgs}$ towards the wall when the Sigma model is used.
For the Vreman model, no decrease can be discerned, although asymptotically $\nu_\mathrm{sgs} \sim y$.
This highlights the fact that it is not only the asymptotic behavior, which is important, but also how fast the model transitions towards the asymptote.
Away from the wall, Vreman and Sigma show similar values of $\nu_\mathrm{sgs}$.

\begin{figure}[htp!]
	\centering
	\includegraphics{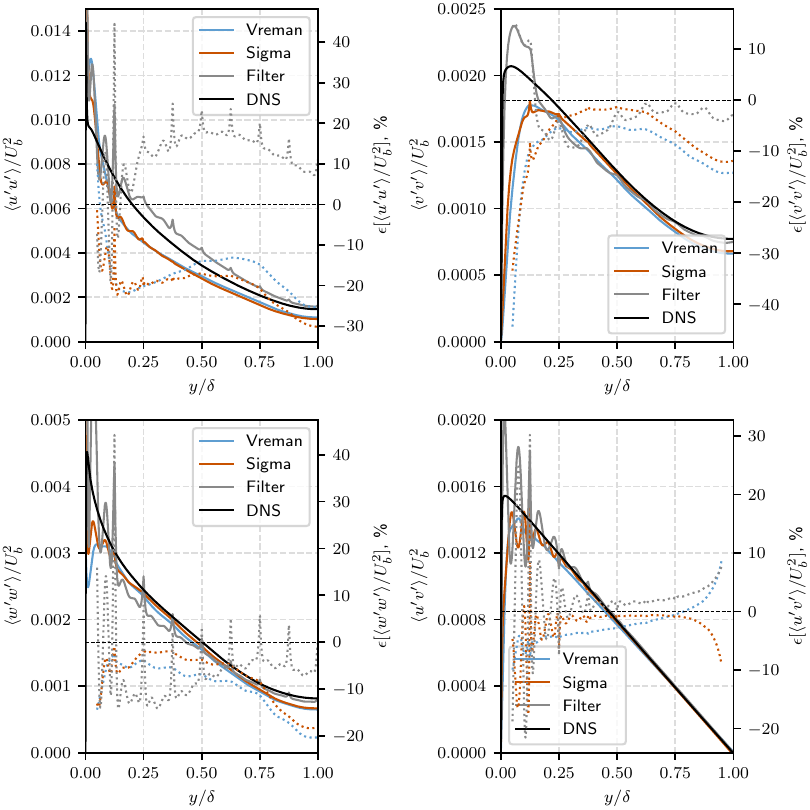}
	\caption{\textit{Solid lines:} Reynolds stresses obtained in channel flow simulations, using the M3 mesh and different SGS models. \textit{Dotted lines:} The relative errors in the profiles with respect to the reference DNS.}
	\label{fig:channel_sgs_re}
\end{figure}

We now turn to the second-order statistics, which are shown in Figure~\ref{fig:channel_sgs_re}.
The profiles obtained with the Sigma and Vreman model do not differ much, except for a somewhat higher level of oscillations close to the wall when using the former.
Using filtering, on the other hand, changes the solution significantly.
Not only do the near-wall oscillations increase, but so does the overall level of velocity variance in the core of the channel.
For $\mean{u'u'}$ this leads to an over-prediction of the DNS data, whereas for $\mean{v'v'}$ and $\mean{w'w'}$ the results are quite accurate, particularly for $y/\delta > 0.75$.
We also see the mesh structure reflected in the results through peaks in the profiles near the boundaries between two elements.
This is similar to the results in Gillyns et al.~\cite{Gillyns2022}.

Finally, we note that the values of $\mean{\tau_w^{wall}}$ change drastically with the SGS model. The error with respect to the DNS is $-25.72$\% when using the Sigma model and $-84.85$\% when using the filtering.
The result for the Vreman model is 8.07\%, see Table~\ref{tab:neumann_tau}.
So, clearly, higher dissipation near the wall benefits the correlation strength between $\tau_w^{wm}$ and $\tau_w^{wall}$.

To summarize, the results in this subsection show that high-pass filtering alone does not introduce enough damping to avoid the contamination of the solution with numerical artifacts.
Both the Vreman and the Sigma model perform well, but the Vreman model could, perhaps, be preferred due to smoother derivatives, as indicated by $\Phi$.

Of course, these results could be altered by changing the model constants.
One could also combine weak filtering with an explicit model, like in~\cite{Chatterjee2017}.
Finally, many other SGS models exist and could be tested, so the parameter space is extremely large.
A general conclusion that can nevertheless be drawn is that in the SEM the model has to provide sufficient dissipation to damp numerical oscillations that can be strong on a coarse grid.
Another SEM-specific challenge is the discontinuity of the derivatives and their spikes near the element boundaries.
Lastly, we saw that there was no significant effect of near-wall $\nu_\mathrm{sgs}$ on the log-layer mismatch, as all three simulations gave accurate $\tau_w^{wm}$ predictions.
This appears to contradict the theory of Brasseur and Wei~\cite{Brasseur2010}, although it should be noted that we only considered one of the three criteria for accurate WMLES that the authors postulate, so a more detailed analysis is warranted.

\subsection{Simulations with the viscosity-based condition}
\begin{figure}[htp!]
	\centering
	\includegraphics{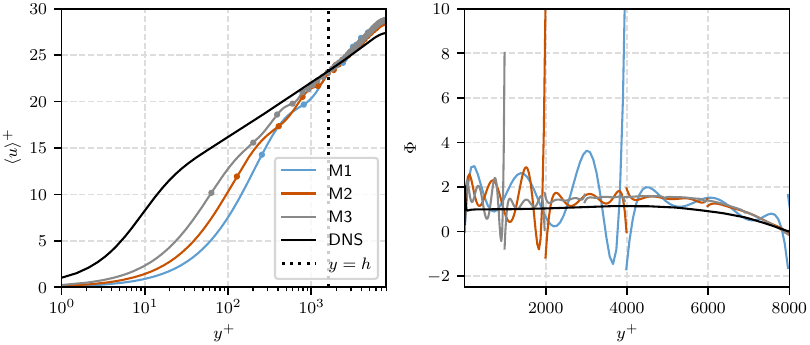}
	\caption{\textit{Left:} Inner-scaled mean velocity profiles obtained in channel flow simulations, using the viscosity-based boundary condition. \textit{Right:} The profiles of the indicator function $\Phi$ from the same simulations.}
	\label{fig:channel_visc_u_inner}
\end{figure}

The simulations using the viscosity-based wall boundary condition (see Section~\ref{sec:nut_condition}) are now considered.
The left plot in Figure~\ref{fig:channel_visc_u_inner} shows the inner-scaled mean velocity profiles.
The no-slip condition leads to a strong under-prediction with respect to the reference profile at $y+ \lessapprox 2500$, reflecting the coupling of velocity values within a single element.
Since the bulk velocity in the channel is fixed, this forces an over-prediction of the velocity in the core of the channel.

The issue propagates also into the values of the velocity derivatives, which are shown in the right plot in Figure~\ref{fig:channel_visc_u_inner} via the indicator function.
A large jump can be seen in $\Phi$ at the edge between the near-wall element and the one adjacent to it.
Note that the jumps are significantly larger than those observed when using the Neumann condition (see Figure~\ref{fig:channel_neumann_u_inner}).
Since SGS models depend on the derivatives, these spikes have a strong affect on the overall momentum balance, as shown below.

\begin{figure}[htp!]
	\centering
	\includegraphics{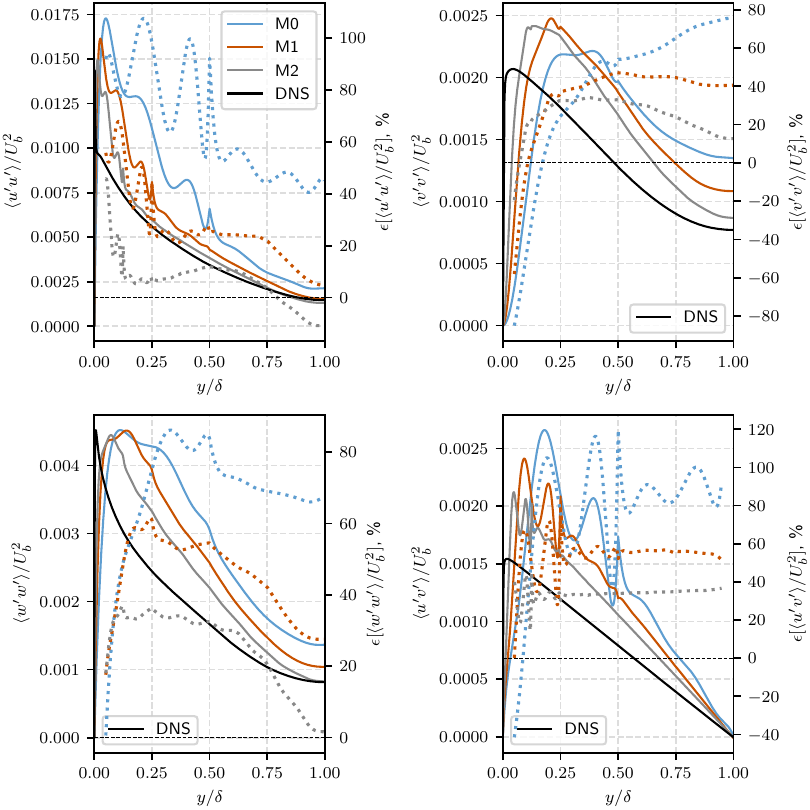}
	\caption{\textit{Solid lines:} Reynolds stress profiles obtained in channel flow simulations, using the viscosity-based boundary condition. \textit{Dotted lines:} The relative errors in the profiles with respect to the reference DNS.}
	\label{fig:channel_visc_re}
\end{figure}

The obtained Reynolds stresses are shown in Figure~\ref{fig:channel_visc_re}.
Unlike the case of the Neumann condition, here a strong over-prediction of all the stresses is observed.
In particular, note that the shear stress is over-predicted in the core of the channel, which can, in principle, only occur due to an error in the wall stress.

To clarify the observed behavior, it is necessary to consider the predicted values of the wall stress, which are provided in Table~\ref{tab:vis_tau}.
What we see is that $\mean{\tau_w^{wm}}$ is actually quite accurate, although an under-prediction is observed, as expected based on the analysis in Section~\ref{sec:nut_condition}.
Furthermore,  $\mean{\tau_w^{wm}}$ and  $\mean{\tau_w^{wall}}$ are now equal, as desired, but both are in strong disagreement with  $\mean{\tau_w^{\nabla p}}$.
The latter is larger than the expected value based on the target $\rey_\tau$.
Analysis shows that it is $\mean{\tau_w^{\nabla p}}$ that correctly scales the Reynolds shear stress and is thus the dynamically relevant quantity.

These results can only be possible when momentum conservation is strongly violated.
To further confirm this, we ran another set of simulations, where the forcing was set explicitly, therefore fixing $\rey_\tau$ instead of $\rey_b$.
In that setup, the velocity profile should adjust itself in such a way that $u^{wm}$ assumes the value that leads the wall model to predict the correct stress, thus balancing out the explicit forcing and fulfilling momentum conservation.
However, this was not observed, and instead the stress is strongly under-predicted.

\begin{table}[htp!]
	\caption{Errors in the values of the wall shear stress obtained in channel flow simulation using the viscosity-based boundary condition.}
	\label{tab:vis_tau}       
        \begin{ruledtabular}
	\begin{tabular}{p{2cm}p{2cm}p{2cm}p{2cm}}
		Mesh & $\epsilon[\mean{\tau_w^{wm}}]$ & $\epsilon[\mean{\tau_w^{wall}}]$ &$\epsilon[\mean{\tau_w^{\nabla p}}]$ \\
		\noalign{\smallskip}\hline\noalign{\smallskip}
		M1 & -0.91\% & -0.91\% &  38.05\% \\
		M2 & -0.85\% & -0.85\% &  25.73\% \\
		M3 & -2.26\% & -2.26\% &  15.65\% \\
	\end{tabular}
        \end{ruledtabular}
\end{table}

Note that the error in $\mean{\tau_w^{\nabla p}}$ decreases with mesh refinement, just like it did for $\mean{\tau_w^{wall}}$ when the Neumann boundary condition was used.
The same behavior is visible in the profiles of $\Phi$.
It thus appears that the same property of the numerical method manifested itself for both types of wall model boundary conditions.
The Neumann boundary condition has the clear advantage that the wall stress value produced by the wall model, $\mean{\tau_w^{wm}}$ agrees with $\mean{\tau_w^{\nabla p}}$ and can therefore be used to compute inner-scaled quantities.

\section{Flat-plate turbulent boundary layer simulations} \label{sec:tbl}

\subsection{Case setup} \label{tbl:setup}
\begin{figure}[htp!]
	\centering
	\includegraphics{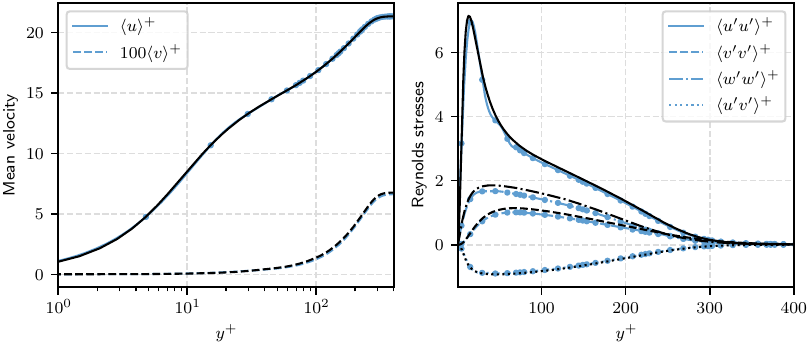}
	\caption{The velocity statistics at the inlet to the flat-plate TBL domain. Black lines show the SIMSON precursor data, and the blue lines the actual statistics obtained in Nek5000, including errors from time averaging and also temporal and spatial interpolation.}
	\label{fig:inflow}
\end{figure}

We now consider a zero-pressure-gradient turbulent boundary layer developing on a flat plate.
The domain is a box of size $450 \times 20 \times 27.395$, measured in $\delta^{in}_{99}$, which is the thickness of the boundary layer at the inlet.
To generate the inflow, plane data has been sampled from a TBL simulation performed in the pseudo-spectral code SIMSON~\cite{Chevalier2007}.
The data was saved at a streamwise location corresponding to $\rey^{in}_\theta = 790$ and interpolated onto the WMLES meshes as a pre-processing step.
Inside Nek5000, the pre-processed planes are read, and cubic temporal interpolation is used to get values at any given time within the range of the dataset, which is $\approx 450 \delta^{in}_{99}/U_0$.
Here $U_0$ is the reference freestream velocity.
Since this corresponds to only a single flow-trough time, the data is reused in cycles.
The profiles of the inflow mean velocity and Reynolds stresses are shown in Figure~\ref{fig:inflow}.

At the top boundary, the flow is allowed to exit the domain, whereas a slip condition is applied for the boundary-parallel velocity components.
At the outlet, the so-called Dong boundary condition~\cite{Dong2014} is used to eliminate possible backflow events, which would deteriorate stability.
Finally, in the spanwise direction, periodic conditions are applied, and at the bottom wall the stress is prescribed using a Neumann condition, coupled with zero wall-normal velocity.

Based on the channel flow results, the Vreman SGS model is used in the simulations.
Spalding's law is still used by the wall model to predict the stress, but here with $\kappa = 0.41$ and $B=5.2$, which is a more classical set of values suitable for lower Reynolds numbers.
The sampling point is set to $0.2\delta_{99}(x)$, with the distribution of the thickness computed using a power-law estimate~\cite{Rezaeiravesh2016}.
An averaging time of $\Delta t_\mathrm{avrg} = 10 \Delta t$ is applied to the wall model's input velocity signal.
It is important to note that for this flow case the accuracy of Spalding's law is lower compared to the tuned version used in the channel flow study.
As demonstrated in Appendix~\ref{sec:appendix_spalding}, a wall-stress error of $\approx \pm 1\%$ is produced when using DNS~\cite{Schlatter2010} and WRLES~\cite{Eitel-Amor2014} data to provide the input velocity value.
For channel flow, the error is an order of magnitude lower.
This could theoretically be mitigated by a priori tuning of $\kappa$ and $B$ and making them a function of the streamwise coordinate (and thus the Reynolds number), but such an approach is unpractical.
Therefore, even for attached TBLs, more accurate wall models that consistently produce correct results across a wide range of Reynolds numbers are desirable.

The channel flow study showed that WMLES requires at least 4 elements across the boundary layer in all three spatial directions to adequately capture second-order statistics.
For the TBL, the meshing strategy should ideally take into account the boundary layer's growth.
Using a structured mesh topology, it is only possible to adjust the element size in the streamwise and wall-normal directions.
In $z$, the size is forced to remain the same, meaning that the outer scales become increasingly better resolved as the TBL grows.
Accordingly, the anisotropy of the mesh also becomes stronger.
For the particular domain length selected here,  $\delta_{99}$ grows by approximately a factor of 8.

\begin{figure}[htp!]
	\centering
	\includegraphics{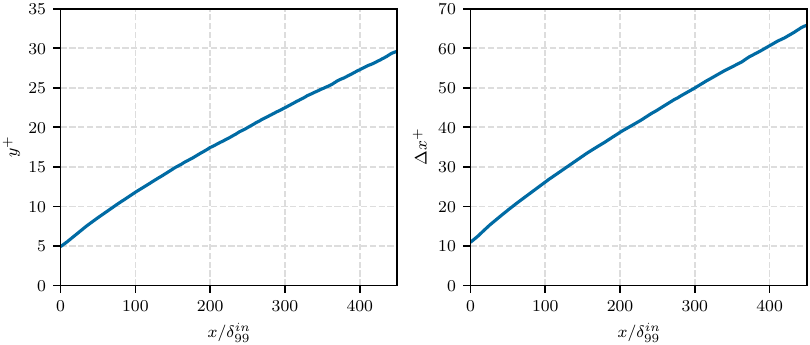}
	\caption{\textit{Left}: The values of $y^+$ at the first off-wall node in the flat-plate TBL simulations. \textit{Right}: The values of $\Delta x^+$ in the flat-plate TBL simulations. The values across each element are averaged to remove the variation due to GLL node distribution. The values of $u_\tau$ used for scaling are taken from the simulation on the M3 mesh.}
	\label{fig:tbl_resolution}
\end{figure}

An alternative is to keep the mesh size constant across all three spatial directions, thus retaining isotropy.
This leads to much larger mesh sizes.
Moreover, it should be considered that at the inflow using 4 elements across $\delta_{99}$ leads to a mesh, which is essentially WRLES in terms of its size in inner units ($\approx 10\delta_\nu$, see also Figure~\ref{fig:tbl_resolution}).
Thus, an insignificant increase in resolution in $y$ could be performed to get a true WRLES mesh, which is well worth removing the additional complications associated with wall modeling.
Therefore, to justify WMLES, the coarsening of the resolution with $x$ and $y$ must be performed.

One could also consider selecting a coarser spanwise resolution at the inlet, at the cost of a larger flow adaptation region.
Here, we perform simulations using 3 meshes: M1, M2, and M3. For each of them, the resolution in $x$ and $y$ is adapted to the growth of the TBL, so that it is covered by approximately 4 elements.
In the spanwise direction, the meshes have, respectively, 1, 2, and 4 elements per $\delta_{99}$ at the inlet.
More details about the mesh sizes are found in Table~\ref{tab:tbl_mesh}.

\begin{table}[htp!]
	\caption{Computational meshes for the flat-plate TBL simulations. $N_x$, $N_y$, and $N_z$ are the number of elements along the three coordinate axes. $N_{gll}$ is the total number of nodes in the mesh.}
	\label{tab:tbl_mesh}
        \begin{ruledtabular}
	\begin{tabular}{p{1cm}p{1cm}p{1cm}p{1cm}p{2cm}}
		Mesh & $N_x$ & $N_y$ & $N_z$ & $N_{gll}$ \\
		\noalign{\smallskip}\hline\noalign{\smallskip}
		M1 & 519 & 15 & 28 & $75 \cdot 10^6$ \\
		M2 & 519 & 15 & 56 & $149 \cdot 10^6$ \\
		M3 & 519  & 15 & 112 & $299 \cdot 10^6$ \\
	\end{tabular}
        \end{ruledtabular}
\end{table}

Figure~\ref{fig:tbl_resolution} shows the resolution of the mesh in inner units. 
As mentioned above, for the M3 mesh the resolution is essentially that of a WRLES at the inlet.
On the two coarser meshes, the wall-parallel resolution balances on the boundary between a very poor WRLES and a pure WMLES.

\subsection{Simulation results}
Before presenting the results, it should be noted that for a developing boundary layer, comparison across different datasets is significantly more difficult than for channel flow.
The main reason are the uncertainties in computing the length scales defining the thickness of the boundary layer.
All three measures, $\delta_{99}$, $\delta^*$ and $\theta$ depend on e.g.~the particular choice of computing the freestream velocity and selecting the upper bound in the integrals defining $\delta^*$ and $\theta$.
Even when comparing published DNS datasets, appreciable differences can be observed, see~\cite{Schlatter2010}.
For WMLES the difficulty is further exacerbated by the fact that the velocity profile in the inner layer is inaccurate, leading to additional errors in $\delta^*$ and $\theta$.

For completeness, we note that here we compute a local $U_0(x)$ as the maximum value of the mean velocity at the current $x$.
Linear interpolation is then used to obtain a $\langle u(y) \rangle$ profile on a very dense equidistant mesh, which is then used to compute $\delta_{99}$.
Finally, $\delta^*$ and $\theta$ are computed using the same profile, with $y|_{ \langle u \rangle =U_0}$ used as the upper bound of the integral. 

Three reference datasets are used to analyze the accuracy of the WMLES results: the DNS of Schlatter and Örlü~\cite{Schlatter2010}, the WRLES of Eitel-Amor et al.~\cite{Eitel-Amor2014}, and the DNS of Sillero et al.~\cite{Sillero2013}.

\begin{figure}[htp!]
	\centering
	\includegraphics{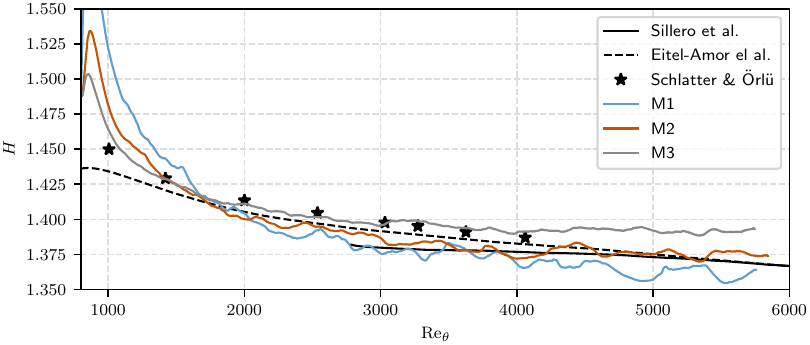}
	\caption{The values of shape factor, $H$, obtained in the flat-plate TBL simulations.}
	\label{fig:shape_factor}
\end{figure}

Similarly to channel flow, we begin with the analysis of velocity statistics in outer scale and integral quantities.
First, the shape factor $H$ is considered.
This quantity is, by definition, highly sensitive to the procedure for computing $\delta^*$ and $\theta$.
This is well-illustrated by the discrepancies observed among the three reference datasets, as seen in Figure~\ref{fig:shape_factor}.
Following adaptation from the non-ideal inflow, all three WMLES curves agree well with the references.
There is no clear advantage of mesh refinement observed.

\begin{figure}[htp!]
	\centering
	\includegraphics{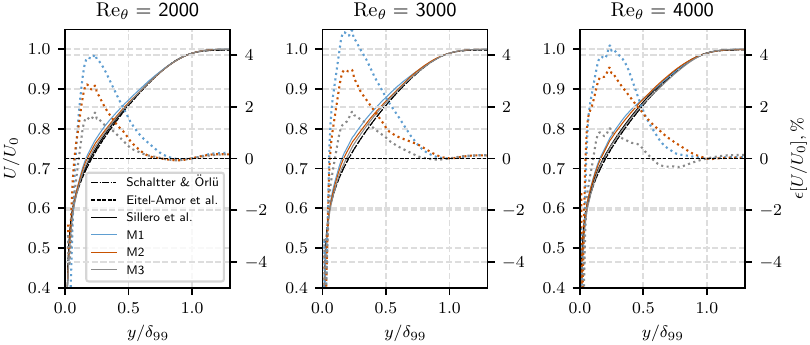}
	\caption{The mean velocity profiles in outer scaling, obtained in the flat-plate TBL simulations. Dotted lines show the relative error profiles with respect to the DNS of Schlatter and Örlü~\cite{Schlatter2010}. At $\rey_\theta = 3000$ the data from Sillero et al.~\cite{Sillero2013} is not available.}
	\label{fig:tbl_u}
\end{figure}

Figure~\ref{fig:tbl_u} shows the mean streamwise velocity profiles at selected $\rey_\theta$.
The WMLES profiles  over-predict the reference data somewhat, and there appears to be a clear trend of better agreement with grid refinement.
At $\rey_\theta = 4000$, the M3 curve is virtually indistinguishable from the references.
The fact that results improve with $\rey_\theta$ is important since it hints that at high Reynolds numbers accuracy on par with channel flow will be achieved.
The reason behind the improvement of the results with mesh refinement in $z$ nevertheless remains an open question, because in the second half of the domain, the outer scales are well resolved in $z$ on all three meshes.
It can be speculated that the decrease $\Delta z^+$ benefits the results, which would violate that premise of WMLES being independent of inner-scale resolution.
While this possibility has to be accepted at the relatively low Reynolds numbers considered here, one should also appreciate the fact that the differences observed among the M1-M3 profiles are small.

It should be noted, that since the $\delta_{99}$ values used to scale the abscissa are also simulation outcomes, the observed over-prediction can to some degree be specific to the chosen scaling.
The consequence of this is that, unlike for channel flow, we cannot, with full confidence, claim that the values sampled at $y / \delta_{99} = h$ are too high and make use of that in the error analysis of the wall modeling.
However, the distribution of $\delta_{99}(x)$ (not shown) is in good agreement with the power-law estimate used to compute it a priori in order to set $h$. 

It is also noteworthy that the relative error profiles peak very close to the chosen $h=0.2\delta_{99}$, and a lower $h$ may have benefited the accuracy of the $\tau_w^{wm}$ predictions, which are discussed below.

\begin{figure}[htp!]
	\centering
	\includegraphics{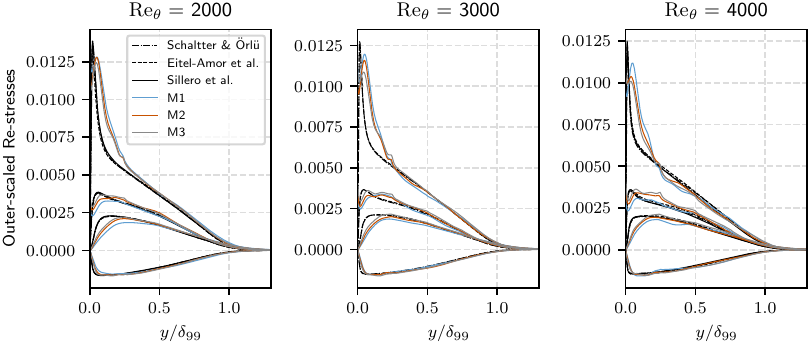}
	\caption{The Reynolds stresses' profiles in outer scaling, obtained in the flat-plate TBL simulations. At $\rey_\theta = 3000$ the data from Sillero et al.~\cite{Sillero2013} is not available.}
	\label{fig:tbl_re}
\end{figure}

Figure~\ref{fig:tbl_re} shows the obtained Reynolds stresses.
Interestingly, the predictions are more accurate than for channel flow.
This may be explained by the lower Reynolds number of the flow, allowing the simulation to partially capture, for example, the peak in $\langle u'u'\rangle$.
Regardless, in the outer layer, the level of accuracy is excellent for a WMLES.

\begin{figure}[htp!]
	\centering
	\includegraphics{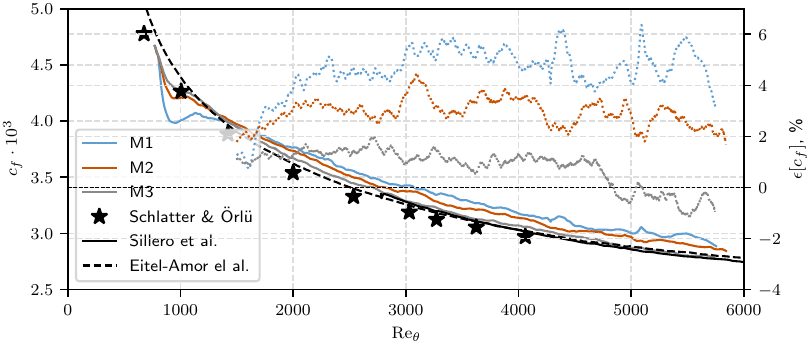}
	\caption{The values of skin friction obtained in the flat-plate TBL simulations, and the corresponding relative errors with respect to the WRLES of Eitel-Amor et al.~\cite{Eitel-Amor2014}.}
	\label{fig:cf}
\end{figure}

Now, the performance of the wall model is evaluated by looking at the skin friction, $c_f$, computed from $\mean{\tau_w^{wm}}$.
The obtained values are shown in Figure~\ref{fig:cf}.
As with the velocity profiles, results improve with mesh refinement.
On the finest mesh, the error is below 2\%, which can be considered the best possible outcome given the errors in Spalding's law discussed in the previous subsection (see also Appendix~\ref{sec:appendix_spalding}).

\begin{figure}[htp!]
	\centering
	\includegraphics{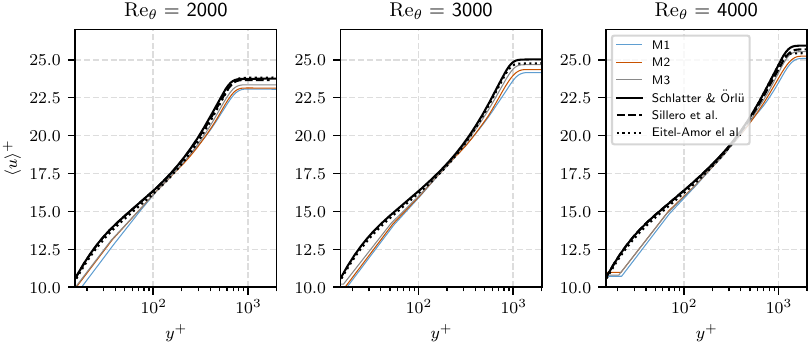}
	\caption{The mean velocity profiles in inner scaling, obtained in the flat-plate TBL simulations. At $\rey_\theta = 3000$ the data from Sillero et al.~\cite{Sillero2013} is not available.}
	\label{fig:tbl_inner}
\end{figure}

The inner-scaled velocity profiles are shown in Figure~\ref{fig:tbl_inner}.
As expected, the over-prediction in $\mean{\tau^{wm}_w}$ leads to under-prediction in $\mean{u}^+$.
At  higher Reynolds numbers, the agreement is generally good, and excellent for the M3 profiles.

In conclusion, we put the presented results into perspective by looking at what has previously been reported in the literature.
Unlike for channel flow, there are quite few WMLES datasets to be found for the flat-plate TBL flow case.
Particularly at higher Reynolds numbers and with rigorous error analysis with respect to a DNS or WRLES.
There is a tendency to use channel flow as the only attached TBL test case and then move on directly to more applied configurations.

Results from WMLES performed in the finite volume solver OpenFOAM are available in~\cite{Mukha2021}.
The accuracy is lower than that demonstrated here, particularly for the Reynolds stresses.
However, the meshes used in those simulations are coarser, corresponding roughly to 2 elements per $\delta_{99}$.
In a recent paper by Lozano-Durán and Bae~\cite{Lozano-Duran2023}, another finite volume solver is used, together with an even coarser mesh. 
The results are rather poor: the velocity profiles deviate from DNS by 10–30\%, and the wall stress between 5\% and 15\%, depending on the wall model.
In~\cite{Bae2019}, the dynamic slip wall model is used, predicting $c_f$ ``well within'' a 4\% error.
The velocity statistics are compared to the DNS of Sillero et al.~\cite{Sillero2013}, with accuracy quite similar to that of the simulations here.

While the overview above is not exhaustive, it can be safely said that the results of the current WMLES are on par with the state of the art, possibly outperforming it for selected quantities, such as $\langle u'u' \rangle$, which is difficult to capture well.

\section{Conclusions} \label{sec:conclusions}
This article presents a detailed study of WMLES methodology using an SEM-based solver, Nek5000.
The principle outcome is that a combination of algebraic wall-stress modelling and explicit SGS modelling using the Vreman model  can be used to successfully conduct WMLES.
The exhibited predictive accuracy is on par with the state of the art, which makes SEM-based WMLES a very attractive approach due to the excellent performance of solvers like Nek5000 on CPUs~\cite{Offermans2016} and, e.g.,~Neko on both CPUs and GPUs~\cite{Jansson2024}.
Indeed, efforts on porting the current Nek5000 implementation to Neko to enable WMLES on GPU-accelerated supercomputers are already under way. 
The possibility to interpolate the solution fields with spectral accuracy to any point is also very beneficial for WMLES, which relies strongly on an accurate input signal for the wall model.
No accuracy issues have been observed when sampling from the wall-adjacent element.

At the same time, our work reveals several difficulties that arise when running SEM simulations on coarse grids.
Global momentum balance is only achieved asymptotically, which led to the attempt of using subgrid viscosity to set the wall stress to be largely unsuccessful.
Similarly, the values of the velocity derivatives at the wall only asymptotically converge to that set by the weakly formulated Neumann boundary conditions.
As a result, for WMLES, the velocity profiles do not exhibit the correct gradient at the wall even when the predictions of the wall model are close to exact.

Directly related to the momentum balance, the velocity derivatives experience jumps at the element boundaries.
Since the SGS models depend on the derivatives, associated non-physical spikes appear in the SGS viscosity values.
Unfortunately, the largest derivative jump occurs at the boundary between the wall-adjacent and its neighbouring element, i.e.~a near-wall region where SGS modelling is particularly difficult.

The SGS model also plays an additional role of numerically stabilizing the simulation.
We show that the Vreman model does an excellent job at this, and we have experienced no issues with stability regardless of simulation settings.
Indeed, explicit SGS modelling is shown here to perform much better in this regard than the otherwise common filter-based approach.
Nevertheless, this is an additional dimension to consider when selecting the SGS model.
In this work we also tested the Sigma model, and it has not been dissipative enough near the wall to avoid very strong wiggles in the velocity derivatives .

Some results of our work are relevant for WMLES in general, regardless of the numerical framework used.
Time-averaging of the wall model's input velocity signal is shown to give only a marginal improvement, if any, when applied to simple algebraic wall models, such as Spalding's law.
Although a reduction of the error has been observed in a priori studies, the effect is not strong enough to systematically improve wall stress predictions in a real simulation.

We also observed no connection between the log-layer mismatch and the shear stress resolution in the first off-wall node.
In spite of the theoretical arguments in~\cite{Brasseur2010}, our solutions exhibit no mismatch although the resolved shear stress is dominated by the SGS one in the direct proximity of the wall.

To conclude, we note that some of the methodological difficulties discussed above can likely be remedied by incorporating additional steps to the solution procedure.
For example, the SGS model could apply smoothing to the velocity derivative values prior to computing the eddy viscosity.
Improving various aspects of SEM-based WMLES remains as a future research direction for the authors.
Another natural continuation of this work is to consider separating flows, which remain a long-standing challenge for WMLES.
Additionally, we are actively working on WMLES of stratified atmospheric boundary layers.

\section{Acknowledgements} \label{sec:acknowledgements}
The computations were enabled by resources provided by the Swedish National Infrastructure for Computing (SNIC) at the PDC Center for High Performance Computing, KTH Royal Institute of Technology, partially funded by the Swedish Research Council through grant agreement no. 2018-05973.
The authors thank Ashwin Vishnu Mohanan, previously at Stockholm University, for providing the implementation of the Vreman model.
The authors thank Simon Sticko, previously at Uppsala University, and Geert Brethouwer, at KTH Royal Institute of Technology, for fruitful scientific discussions, which helped facilitate the investigations reported in this article. 


\appendix
\section{Spalding's law accuracy analysis}
\label{sec:appendix_spalding}
In this section we consider the accuracy of Spalding's law as a wall model, which of course directly relates to how well it approximates the mean velocity profile.
The accuracy is quantified using mean velocity profiles from DNS data.
In other words, we ask the question: If the input to the wall model would be the exact mean velocity value, what error in $\tau_w$ can we expect due to the inaccuracy in the wall model's governing equation?

\begin{figure}[htp!]
	\centering
	\includegraphics{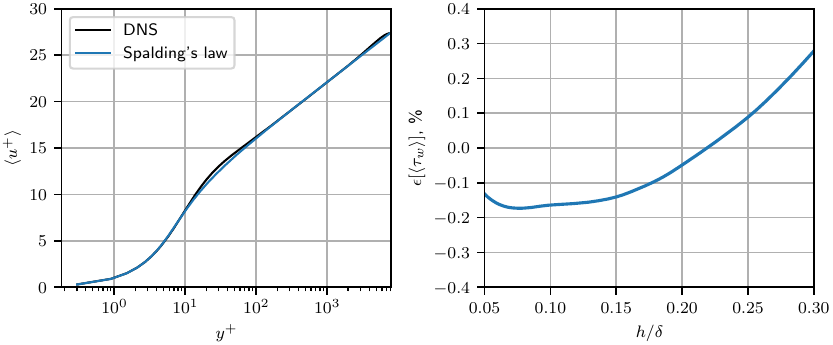}
	\caption{\textit{Left:} Channel flow mean velocity profile at $\rey_\tau \approx 8000$ and the approximation by Spalding's law. \textit{Right}: The relative error in $\mean{\tau_w}$ produced by Spalding's law with DNS mean velocity data as input.}
	\label{fig:spalding_channel}
\end{figure}

Channel flow at $\rey_\tau \approx 8000$ is considered first.
The parameters $\kappa=0.387$, $B=4.2$ are used for Spalding's law, following~\cite{Yamamoto2018}.
The left plot in Figure~\ref{fig:spalding_channel} shows the law along with the reference DNS profile.
The approximation is clearly very accurate, excluding the buffer region, where the law under-predicts the reference profile somewhat.
Examination reveals that in the log-law region the law's relative error with respect to the DNS does not exceed 0.2\%, which reflects the excellent tuning of $\kappa$ and $B$.

In the right plot in the same figure, the relative error in $\mean{\tau_w}$ is shown as a function of the sampling height, $h/\delta$.
As expected, the errors are small and remain within the $[-0.2\%, 0.2\%]$ band for $h/\delta \in [0.05 ,\approx0.25]$.
At $h/\delta = 0.2$ (which is used here in the channel flow WMLES) the model gives slight under-prediction of about 0.05\%.

\begin{figure}[htp!]
	\centering
	\includegraphics{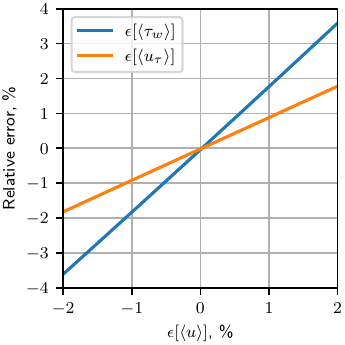}
	\caption{The propagation of error in $\langle u \rangle$ into $\mean{u_\tau}$ and  $\mean{\tau_w}$.}
	\label{fig:spalding_channel_propagation}
\end{figure}

It is also interesting to see how the law propagates the errors in the sampled velocity to the values of the stress.
This relationship is shown in Figure~\ref{fig:spalding_channel_propagation}, which is obtained by perturbing the DNS value of $\mean{u}$ prior to using it as input.
For both $\mean{u_\tau}$ and $\mean{\tau_w}$ it remains essentially linear for relevant values of the error in $\mean{u}$.
The difference is in the slope of the line, which is twice as large for the case of $\mean{\tau_w}$.
As a rule of thumb, one can say that error in the stress is about double the error in velocity.

\begin{figure}[htp!]
	\centering
	\includegraphics{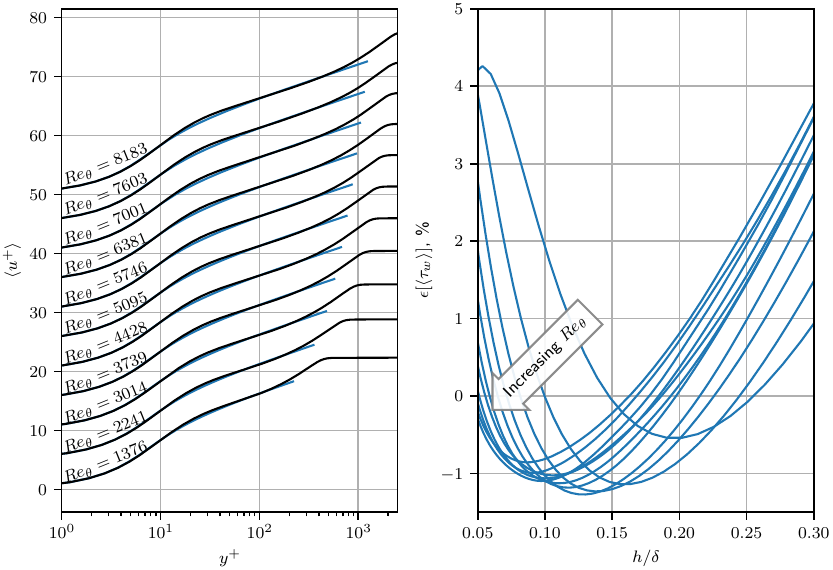}
	\caption{\textit{Left:} Flat-plate TBL mean velocity profiles at increasing $\rey_\theta$ and the approximation by Spalding's law. Successive profiles shifted by 5 inner units along the ordinate. \textit{Right}: The relative error in $\mean{\tau_w}$ produced by Spalding's law with DNS mean velocity data as input.}
	\label{fig:spalding_tbl}
\end{figure}

A similar analysis is now performed for the flat-plate TBL.
To that end, we consider wall-resolved LES data spanning $\rey_\theta$ values between $\approx 1376$ and $\approx 8183$~\cite{Eitel-Amor2014} as reference.
The velocity profiles together with Spalding's law are shown in the left plot in Figure~\ref{fig:spalding_tbl}.
Here, we use $\kappa = 0.41$ and $B = 5.5$.
There is reasonable agreement in the log-law region, but some issues have to be highlighted.

First, it should be noted that Spalding's law approaches its limiting logarithmic behaviour around $y^+ = 150$.
At low Re numbers, this is very close to the wake region, meaning that for reasonable values of $h/\delta$, one would be in law's buffer region, where accuracy is subpar, and calibration via log-law constants is difficult.
Second, it should be appreciated that at all Re numbers the agreement here is significantly worse than in the channel flow case above, which had a long log-layer and highly tuned model constants.
Both of these issues could be brushed off as ``low Re number effects'' and thus not relevant for WMLES.
However, for several interesting flow cases (e.g.~exhibiting separation) benchmark data is available only at relatively low Re number; $\rey_\theta \approx 1000$ is not uncommon.

The observations above find their reflection in the curves in the right plot of Figure~\ref{fig:spalding_tbl}, showing the obtained error curves in $\mean{\tau_w}$.
At low Re numbers we observe higher errors at lower $h/\delta$.
As the Re grows, the error profile stabilizes, but the values remain about an order of magnitude higher compared to the channel flow case.

\begin{figure}[htp!]
	\centering
	\includegraphics{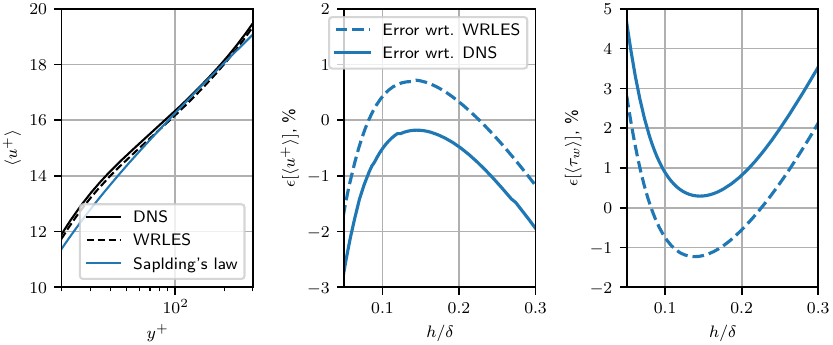}
	\caption{\textit{Left:} Flat-plate TBL mean velocity profiles at $\rey_\theta \approx 3000$ from DNS~\cite{Schlatter2010} and WRLES~\cite{Eitel-Amor2014}; the approximation by Spalding's law.
	\textit{Middle:} The relative error in Spalding's law with respect to the DNS and WRLES.
    \textit{Right}: The relative error in $\mean{\tau_w}$ produced by Spalding's law with DNS and WRLES mean velocity data as input.}
	\label{fig:spalding_tbl2}
\end{figure}

The fact that the error profiles cross 0 may tempt to carefully match $h$ to exactly those values where the crossing occurs.
However, this location is very sensitive to what one considers the ``ground truth''.
To illustrate this, we consider both the wall-resolved LES data used before, but also DNS from~\cite{Schlatter2010}, the former at $\rey_\theta \approx 3014$ and the latter at $\rey_\theta \approx 3030$.
Both profiles are shown in the left plot of Figure~\ref{fig:spalding_tbl2}.
The DNS profile lies a little bit above the WRLES one, which is reflected in the error profile of Spalding law's mean velocity prediction (middle plot in the figure).
Clearly, the law crosses the WRLES profile, but remains below the DNS one.
As consequence, we don't observe a crossing of the zero-error line in the prediction of the wall stress in the case of the DNS.

\bibliography{library}

\end{document}